\documentclass{bioinfo}
\copyrightyear{2015}
\pubyear{2015}

\newcommand{\change}[1]{#1}

\usepackage{listings}
\begin{document}
\firstpage{1}

\title[Rust-Bio]{Rust-Bio---a fast and safe bioinformatics library}
\author[K\"oster]{Johannes K\"oster}
\address{Center for Functional Cancer Epigenetics, Dana-Farber Cancer Institute\\
Department of Biostatistics and Computational Biology, Dana-Farber Cancer Institute, Harvard School of Public Health\\
Department of Medical Oncology, Dana-Farber Cancer Institute, Harvard Medical School}

\history{Received on XXXXX; revised on XXXXX; accepted on XXXXX}

\editor{Associate Editor: XXXXXXX}

\maketitle

\begin{abstract}
\section{Summary:} We present Rust-Bio, the first general purpose bio\-informatics library for the \change{innovative} \emph{Rust} programming language.
Rust-Bio leverages the unique combination of speed, memory safety and high-level syntax offered by Rust to provide a fast and safe set of bioinformatics algorithms and data structures with a focus on sequence analysis.

\section{Availability:}
Rust-Bio is available open-source under the MIT license at \href{https://rust-bio.github.io}{https://rust-bio.github.io}.

\section{Contact:} \href{koester@jimmy.harvard.edu}{koester@jimmy.harvard.edu}
\end{abstract}

\section{Introduction}

With ever increasing amounts of experimental data being generated, their computational analysis becomes \change{increasingly} challenging.
For novel or custom problems where carefully engineered high-performance standalone tools (like read mappers) are not yet available, general purpose bioinformatics libraries can help to minimize the coding effort.
Bioinformatics libraries are published for many popular programming languages, e.g., SeqAn for C++, Biopython, Bioperl and BioRuby \citep{Doring2008,Cock2009,Stajich2002,Goto2010}.
Choosing the programming language for a specific task usually entails a tradeoff between execution and development speed.
Low-level system programming languages like C or C++ provide optimal performance at the cost of increased complexity.
Higher level languages like Python or Perl \change{provide a more concise} syntax while leading to computational overhead introduced by online memory management (e.g. reference counting or garbage collection), type inference and not being compiled but interpreted during execution.
Often, the combination of a high-level language with some carefully engineered implementations of a bioinformatics library is a good choice to quickly solve a problem with reasonable performance.
\change{However, the amounts of data the bioinformatics community is facing in the coming years and the need to handle nature's resources carefully implies that using a high-performance, compiled language is still beneficial for certain problems.}

Recently, \emph{Rust}\footnote{\href{http://www.rust-lang.org}{http://www.rust-lang.org}} has gained attention as a new programming language combining speed with memory safety and high-level syntactical features.
Being compiled with LLVM \citep{Lattner2004}, Rust has many advantages of low-level, system programming languages, such as speed and a small memory footprint.
Supporting automatic type inference, it's code is often less verbose than C or C++ code.
With Rust, type inference happens at compile time, such that runtime overhead (appearing with scripting languages like Python) can be avoided.
The key feature of Rust is a concept of ownership and borrowing of variables, that enables the compiler to automatically decide about lifetime of objects during compile time, making an online memory management superfluous without requiring manual freeing of resources.
At the same time, this concept prevents common sources of errors with low-level languages like accessing invalid memory regions.
\change{Finally, the ownership concept enforces thread-safety, such that race conditions cannot occur.}
These features make Rust a promising solution to above tradeoff problem.

In this work, we present Rust-Bio, the first general purpose bioinformatics library for the Rust programming language.
Rust-Bio provides a high-level, fast and safe API for many state-of-the-art data structures and algorithms used in bioinformatics.

\section{Library}

Rust-Bio is built with the following principles in mind.
Where possible, iterators are returned. This allows to process streams of data with minimal memory footprint. On top, using the extensive set of iterator tools available in Rust, iterators can be e.g. filtered, modified, chained or combined in an easy way.
If a language data type appears suitable, we avoid to enclose data into a custom object. This mimimizes memory usage and increases flexibility when handling the data: e.g. biological sequences are represented as vectors or slices of bytes in ASCII encoding. This allows to use sequences with all algorithms and functions in e.g. the Rust standard library that work with byte vectors or slices.
Each implemented algorithm is automatically tested via continuous integration\footnote{\href{https://travis-ci.org}{https://travis-ci.org}}.
For each algorithm and data structure, we provide complexities in the documentation. Where more than one alternative is available, the documentation tries to highlight distinguishing use cases.
So far, Rust-Bio is focused on algorithms and data structures for biological sequences.
A central component of Rust-Bio are \emph{alphabets}, which, e.g., allow to check in linear time whether a given sequence is a word over the alphabet, transform symbols to their lexicographical ranks and perform bit-encoding to save memory or iterate over q-grams.
Rust-Bio can read and write common file formats like FASTA, FASTQ and BED.  For SAM/BAM, CRAM, and VCF/BCF support it is complemented by Rust-HTSlib.

Especially when considering sequencing data, many problems can be solved with a set of well established data structures like suffix arrays \citep{Manber1990}, the Burrows-Wheeler Transform \citep{Burrows1994}, rank/select data structures \citep{Jacobson1988} and $q$-gram indices.
In line with that, Rust-Bio implements induced sorting for suffix array construction \citep{Nong2009}, the FM-Index \citep{Ferragina2000} for  pattern matching on top of the Burrows-Wheeler Transform, a practical variant of a rank/select data structure \citep{Gonzalez2005} and a $q$-gram index for arbitrary alphabets and $q \leq 32$.
Further, Rust-Bio implements the FMD-Index \citep{Li2012}, that allows to find supermaximal exact matches in DNA sequences and their reverse complements in linear time.

Implementations for many classical pattern matching algorithms are provided, including the algorithm of Knuth, Morris and Pratt, Backward Nondeterministic DAWG Matching, Backward Oracle Matching, the algorithm of Horspool, and the Shift-And algorithm \citep{Knuth1977,GonzaloNavarro,Allauzen1999,Horspool1980,Wu1992}.
\change{In the supplement, we compare the speed of these algorithms against the C++ based Seqan, which is among the fastest bioinformatics libraries \citep{Doring2008}. The benchmarks exemplify that the speed of Rust-Bio is comparable to that of C++ based implementations.
For approximate pattern matching}, Ukkonen's dynamic programming based algorithm \citep{Ukkonen1985} and Myer's bit-parallel algorithm \citep{Myers1999} are provided.
Finally, Rust-Bio implements local, global and semi-global pairwise sequence alignment as variants of the Smith-Waterman and Needleman-Wunsch algorithms \citep{Needleman1970, Smith1981}.
An example for using the Rust-Bio API can be seen in Listing \ref{lst_example}.

\begin{lstlisting}[caption={Creating an FM-Index for a given text with an occurence table sampling rate of 3. Here, the alphabet is used to provide guarantees for being able to limit memory usage during FM-Index construction. Afterwards, we iterate over a FASTQ file, use the alphabet to validate read sequences and search for exact matches in the FM-Index.},float=t,label=lst_example,belowskip=-1.5\baselineskip]
let alphabet = alphabets::dna::iupac_alphabet();
let pos = suffix_array(text);
let bwt = bwt(text, &pos);
let fmindex = FMIndex::new(&bwt, 3, &alphabet);

let reader = fastq::Reader::from_file("reads.fastq");
for record in reader.records() {
    let seq = record.seq();
    if alphabet.is_word(seq) {
        let interval = fmindex.backward_search(seq.iter());
	let positions = interval.occ(&pos);
    }
}
\end{lstlisting}

\section{Conclusion}
Rust-Bio is a general purpose bioinformatics library.
Building on the innovative Rust programming language, Rust-Bio combines memory safety with speed, complemented by rigorous continuous integration tests.
So far, a wide set of algorithms and data structures for biological sequences is provided, ranging from index data structures to pattern matching and alignment, complemented by readers and writers for common file formats.

\section*{Acknowledgements}
We thank Christopher Schr\"oder and Peer Aramillo Irizar for code contributions as well as Sven Rahmann, Dominik Kopczynski, Tobias Marschall and Marcel Martin for their inspiring lecture notes "Algorithms on Sequences".

\bibliographystyle{natbib}
\bibliography{Mendeley}

\end{document}

\section*{Rust-Bio---a fast and safe bioinformatics library: Supplement}
\subsection*{Benchmarks of pattern matching algorithms}
Since Rust-Bio is based on a compiled language, similar performance to C/C++ based libraries can be expected. Indeed, we find the pattern matching algorithms of Rust-Bio to perform in the range of the C++ library SeqAn\footnote{\url{http://www.seqan.de}}:

\begin{center}
\begin{tabular}{lrr}
\hline
Algorithm & Rust-Bio & Seqan \\
\hline
BNDM  &	77ms &	80ms \\
Horspool &	122ms &	125ms \\
BOM &	103ms &	107ms \\
Shift-And &	241ms &	545ms \\
\hline
\end{tabular}
\end{center}
We measured $10,000$ iterations of searching pattern
$$GCGCGTACACACCGCCCG$$
in the sequence of the human MT chromosome (assembly hg38). Initialization time of each algorithm for the given pattern was included in each iteration. Benchmarks were conducted with \emph{Cargo bench} for Rust-Bio and \emph{Python timeit} for SeqAn on an Intel Core i5-3427U CPU. Benchmarking SeqAn from Python timeit entails an overhead of around 1.46ms for calling a C++ binary.
This overhead was subtracted from above run times.

Note that this benchmark only compares the two libraries to exemplify that Rust-Bio has comparable speed to C++ libraries: all used algorithms have advantages depending on text and pattern structures and lengths. Details about when to use which pattern matching algorithm can be found in the documentation of Rust-Bio's pattern matching module.